\documentclass[10pt]{iopart}

\usepackage{graphicx}
\usepackage{color}
\usepackage{lineno}

\newcommand{\J}{\mathrm{J}}
\newcommand{\Kg}{\mathrm{Kg}}

\newcommand{\Celsius}{\,^{\circ}\mathrm{C}}

\newcommand{\m}{\mathrm{m}}
\newcommand{\s}{\mathrm{s}}
\newcommand{\W}{\mathrm{W}}

\begin{document}

\title[Thermal characteristics of LAB]{Thermal diffusivity and specific heat capacity of linear alkylbenzene}

\author{Wenjie Wu$^{1,2}$\footnote[1]{These authors contributed equally to this work}, Guolei Zhu$^3$\footnotemark[1], Qingmin Zhang$^4$, Xiang Zhou$^{1}$\footnote{Corresponding author}, Yayun Ding$^5$, Haoxue Qiao$^1$ and Jun Cao$^5$}
\address{$^1$ Hubei Nuclear Solid Physics Key Laboratory, Key Laboratory of Artificial Micro- and Nano-structures of Ministry of Education, and School of Physics and Technology, Wuhan University, Wuhan 430072, China}
\address{$^2$ Key Laboratory of Particle \& Radiation Imaging (Tsinghua University), Ministry of Education, Tsinghua University, Beijing 100084, China}
\address{$^3$ School of Marine Science and Technology, Northwestern Polytechnical University, Xi’an 710072, China}
\address{$^4$ Department of Nuclear Science and Technology, School of Energy and Power Engineering, Xi'an Jiaotong University, Xi'an 710049, China}
\address{$^5$ Institute of High Energy Physics, Chinese Academy of Sciences, Beijing 100049, China}
\ead{\mailto{xiangzhou@whu.edu.cn}}

\begin{abstract} 
We report the measurements of the thermal diffusivity and the isobaric specific heat capacity of linear alkylbenzene at about 23$\Celsius$ with the standard atmosphere, which are measured for the first time. The conductivity, heat capacity ratio, and speed of sound were derived from our measurements based on thermodynamic relations. The temperature dependence of heat capacity ratio and isobaric specific heat capacity were calculated and relevant results were discussed.
\end{abstract}

%
\vspace{2pc}

\submitto{\PS}
%
%
\ioptwocol
%

\section{Introduction}
Liquid scintillator (LS) is applied to neutrino detection since the first observation of neutrino in 1956~\cite{cowanDetectionFreeNeutrino1956}. It plays an important role in various neutrino researches such as the solar neutrino problem~\cite{kamlandcollaborationFirstResultsKamLAND2003}, precise measurements of the third mixing angle $\theta_{13}$~\cite{anObservationElectronAntineutrinoDisappearance2012,renocollaborationObservationReactorElectron2012,doublechoozcollaborationIndicationReactorOverline2012}, and the detection of neutrino-less double beta decay~\cite{andringaCurrentStatusFuture2016}. In order to collect enough signals, the fiducial mass of the neutrino target should be as large as possible, since neutrinos are interacting with matter via the weak interaction only. Meanwhile, the radioactive backgrounds of LS are required to be as low as possible to gain sensitivity. For a monolithic gigantic detector, the thermal equilibrium condition for LS is difficult to achieve. A nonuniform distribution of temperature could cause the convection of LS inside the detector. The convection is an obstacle to the detection of low energy and rare events since it would bring peripheral backgrounds, such as $^{210}$Pb, to the central volume of the detector. 

Recently, the non-equilibrium phenomenon is observed by Borexino, which appeares as an annual modulation in the spatial distribution of $^{210}$Po inside the detector filled with about 280 tonnes LS~\cite{bravo-bergunoBorexinoThermalMonitoring2018}. The correlation between environmental temperature and background distribution indicates that the migration of backgrounds is driven by the fluid transportation due to thermal convection. One of the main goals of the last operation phase of Borexino is the observation of the CNO solar neutrinos whose dominant radioactive background is $^{210}$Bi. The purified LS after phase one and the thermal insulation of the detector give a possibility to achieve this goal. Precise determination of $^{210}$Bi concentration is essential to the sensitivity of CNO neutrinos. The $\beta^{-}$ decay of bismuth is indistinguishable from the $\nu_{e}$-$e^{-}$ elastic scattering. Conversely, the $\alpha$ decay of polonium can be effectively picked out through Pulse-Shape Discrimination techniques. Thus the concentration of $^{210}$Bi can be determined through the selection of $^{210}$Po while the secular equilibrium is reached. A stable condition of $^{210}$Po is of importance to the high precision $^{210}$Bi concentration. Many efforts have been devoted to monitor and control the thermal environment of Borexino~\cite{bravo-bergunoBorexinoThermalMonitoring2018}. Other LS detectors could also be suffered from the stability problem of backgrounds caused by thermal convection. Therefore, the thermal quantities of LS are necessary to be measured for studying the performance of the detector.

LS usually consists of a solvent and a scintillating solute. Linear alkylbenzene (LAB) is one kind of solvent which is used in Daya Bay, RENO and SNO+~\cite{anObservationElectronAntineutrinoDisappearance2012,renocollaborationObservationReactorElectron2012,andringaCurrentStatusFuture2016,fischerSearchNeutrinolessDoublebeta2018}. It is taken into consideration for the next generation neutrino experiments as well, such as JUNO, Jinping Neutrino Experiment and LENA~\cite{anNeutrinoPhysicsJUNO2016,beacomPhysicsProspectsJinping2017,wurmNextgenerationLiquidscintillatorNeutrino2012a}.
LAB is a mixture of organic compounds with the formula of $\mathrm{C_6H_5C}_n\mathrm{H}_{2n+1}~(n=10\sim 13)$~\cite{dingNewGadoliniumloadedLiquid2008}. Alkylbenzenes with different numbers of alkyls have varied properties. Thus LABs in two batches could have different performance. For constructing a gigantic detector, a large quantity of LAB is required. The temperature difference between the experimental hall and the detector would cause a large amount of heat transfer. Hence the heat capacity of LAB is needed to estimate the requirements to the filling and cooling processes for controlling the temperature of the detector. In addition, the hydrodynamic motion inside the detector should be well-studied for possible physics analysis. The thermal diffusivity is one of the input parameters to do the computational fluid dynamics simulation.

In this paper, we report the measurements of the thermal diffusivity and the isobaric specific heat capacity of LAB at about 23$\Celsius$ with 1 atm. We have deduced the thermal conductivity, heat capacity ratio and speed of sound based on these measurements.

\section{Experimental method}
\subsection{The thermal diffusivity}
The thermal diffusivity $\alpha$ was measured by a dynamic light scattering (DLC) method, and the schematic diagram can be found in literature~\cite{wangThermalDiffusivityDiisopropyl2014}. The light emitted from a laser was split in two. The more intense one was guided to transmit the sample and the weaker one was used as a reference to make the apparatus working at the heterodyne mode. Mixed light of the scattered light and the reference light was limited by two pin holes to make sure that only light scattered by a small volume was detected by the photon counting head. The time autocorrelation function $G^{2}(\tau)$ was measured by a digital correlator in the cross correlation mode. Based on a least-squares fitting of $G^{2}(\tau)$, the decay constant $\tau_{R}$ can be determined and the thermal diffusivity was calculated by $\alpha=1/(q^{2}\tau_{R})$, where $q$ is the modulus of scattering vector which can be derived from the wavelength and incident angle of incident light~\cite{wangThermalDiffusivityDiisopropyl2014}. The pressure of the sample was controlled by a pump system and was monitored by a pressure transmitter (Rosemont 3051S). An electric heater was deployed in the sample container and a platinum resistance thermometer was used to measure the temperature. The relative uncertainties are $3.76\times 10^{-5}$, 0.001, 0.015 for the measurements of wavelength, incident angle and the decay constant respectively. The relative uncertainty of thermal diffusivity is $\pm$1.5\% which was derived through error propagation~\cite{wangThermalDiffusivityDiisopropyl2014}.

\subsection{The isobaric specific heat capacity}
The isobaric specific heat capacity $c_{P}$ was measured by a flow calorimeter, and the schematic diagram can be found in literature~\cite{heMeasurementIsobaricHeat2015}. The apparatus consists of a plunger type pump, a preheater, a calorimeter and a container for the measured sample. The pump (Scientific Systems, Series 1500 HPLC Pump) and the preheater were used to provide samples with controlled flow rate, pressure and temperature. A micro-heater was deployed inside the calorimeter as a heat source. The temperature increment of the sample entering and exiting the calorimeter $\Delta T$ was measured by two platinum resistance thermometers (PRT, Fluke Corporation). Due to the inefficiency of the heat absorption of the sample, a fraction of the heat was absorbed by the calorimeter. The sample collected by the container was weighed to calibrate the flow rate with an analytical balance (ME204, Mettler Toledo). Then the isobaric heat capacity can be determined by the relative difference of the thermal power of the micro-heater with respect to the variance of the flow rate once $\Delta T$ is fixed. High accuracy was reached by keeping the calorimeter adiabatic from the environment. The experimental uncertainties are $\pm$0.0001~W, $\pm$0.001~g/s and $\pm$0.01~K for the measurement of thermal power, flow rate and temperature respectively. The relative uncertainty of isobaric heat capacity is $\pm$0.98\% which was derived through error propagation~\cite{heMeasurementIsobaricHeat2015}.

\section{Results}
\subsection{Measurement results}
The thermal diffusivity $\alpha$ was measured by the DLC method at 22.98$\Celsius$ with 1 atm to be
\begin{equation}
    \alpha=(7.245 \pm 0.109)\times10^{-8}~\m^{2}\cdot\s^{-1}.
\end{equation}

The isobaric specific heat capacity $c_{P}$ was measured by the flow method at 23.15$\Celsius$ with 1 atm to be
\begin{equation}
    c_{P}=2304\pm 23
    ~\J\cdot\Kg^{-1}\cdot\Celsius^{-1}.
\end{equation}

\subsection{Thermal conductivity, isochoric specific heat capacity, heat capacity ratio and thermodynamic speed of sound}
Thermal conductivity $\lambda$ is the property of a material to transfer heat， which is defined as
\begin{eqnarray}
    \lambda =\alpha\rho c_{P}=(0.1426 \pm 0.0026
    )~\W\cdot\m^{-1}\cdot\Celsius^{-1},
\end{eqnarray}
 where $\rho$ is the density of LAB which has been measured at different temperatures and pressures~\cite{zhouDensitiesIsobaricThermal2015}. 
 
 The isochoric specific heat capacity can be obtained based on thermodynamic relations as~\cite{landauStatisticalPhysicsThird1980}
 \begin{equation}
 \label{eq:heatCapacity}
    c_{V}=c_{P}-\frac{T\beta_{P}^{2}}{\rho\kappa_{T}}=1950 \pm 24
    ~\J\cdot\Kg^{-1}\cdot\Celsius^{-1},
 \end{equation}
 where $\beta_{P}$ is the isobaric thermal expansion coefficient and $\kappa_{T}$ is the isothermal compressibility. $\beta_{P}$ and $\kappa_{T}$ were analyzed based on the variance of density with respect to temperature or pressure~\cite{zhouDensitiesIsobaricThermal2015}. 
 
The heat capacity ratio $\gamma$ for LAB is defined as
\begin{equation}
\label{eq:gamma}
    \gamma=\frac{c_{P}}{c_{V}}=1.182 \pm 0.005.
\end{equation}
The uncertainty of $\gamma$ is dominated by the measurements of $\beta_{P},~\rho$ and $\kappa_{T}$ since the correlated uncertainty from $c_{P}$ are canceled. The relative uncertainty of $\gamma$ reaches 0.4\%.

The adiabatic compressibility $\kappa_{S}$ can be derived as
\begin{equation}
\label{eq:kappa_s}
    \kappa_{S}=\frac{\kappa_{T}}{\gamma}=(6.553\pm 0.039
    )\times 10^{-4}~\mathrm{MPa}^{-1},
\end{equation}
where $\kappa_{T}$ is the isothermal compressibility which has been measured in the literature~\cite{zhouDensitiesIsobaricThermal2015}. 

According to the Newton-Laplace formula, the thermodynamic speed of sound in LAB can be calculated as
\begin{equation}
\label{eq:measuredVSound}
    v_{\mathrm{sound}}=\frac{1}{\sqrt{\kappa_{S}\rho}}=1336.34\pm 3.97
    ~\m\cdot\s^{-1}.
\end{equation}

\subsection{Estimation of the temperature dependence of $\gamma$ and $c_{P}$}
Different experiments may work at different temperatures. The temperature dependence of $\gamma$ and $c_{P}$ can be useful. According to a previous research~\cite{zhouDensitiesIsobaricThermal2015}, the density at different temperatures with 0.1 MPa was described as
\begin{equation}
\label{eq:temperatureDependenceRho}
    \rho(t)=\rho_{0}[1-\beta_{0}(t-t_{0})],
\end{equation}
where $t$ is the celsius temperature, $t_{0}$ is 23$\Celsius$, $\rho_{0}$ is the density at $t_{0}$ with 0.1 MPa. And the isobaric thermal expansion coefficient at different temperatures was described as
\begin{equation}
\label{eq:temperatureDependenceBeta}
    \beta_{P}(t)=\frac{\beta_{0}}{1-\beta_{0}(t-t_{0})},
\end{equation}
where $\beta_{0}$ is the isobaric expansion coefficient at $t_{0}$ with 0.1 MPa. The isothermal compressibility at different temperatures were fitted with an empirical equation
\begin{equation}
\label{eq:temperatureDependenceKappa}
    \kappa_{T}(t)=\kappa_{0}+\kappa_t^{\prime}(t-t_{0}),
\end{equation}
where $\kappa_{0}$ is the isothermal compressibility at $t_{0}$, and $\kappa_t^{\prime}$ is an empirical coefficient. The speed of sound at different temperatures was described by a linear function as
\begin{equation}
\label{eq:temperatureDependenceVSound}
    v_{\mathrm{sound}}(t)=v_{0}+v_t^{\prime}(t-t_{0}),
\end{equation}
where $v_{0}$ is the speed of sound at $t_{0}$ and $v_t^{\prime}$ is the gradient of speed of sound with respect to the temperature.

The heat capacity ratio at 0.1 MPa can be derived as
\begin{equation}
\label{eq:temperatureDependenceGamma}
    \gamma(t)=\rho(t)\kappa_{T}(t)v^{2}_{\mathrm{sound}}(t)
\end{equation}
from \eref{eq:kappa_s}\eref{eq:measuredVSound} while $\rho(t)$, $\kappa_{T}(t)$ and $v_{\mathrm{sound}}(t)$ are known. 
Similarly, the isobaric specific heat capacity at 0.1 MPa can be derived as 
\begin{equation}
\label{eq:temperatureDependenceHeatCapacity}
    c_{P}(t)= \frac{v^{2}_{\mathrm{sound}}(t)\beta^{2}_{P}(t)t}{\rho(t)\kappa_{T}(t)v^{2}_{\mathrm{sound}}(t)-1}
\end{equation}
from \eref{eq:heatCapacity}\eref{eq:gamma}\eref{eq:temperatureDependenceGamma} while $\rho(t)$, $\beta_{P}(t)$, $\kappa_{T}(t)$ and $v_{\mathrm{sound}}(t)$ are known.
$\rho(t)$, $\beta_{P}(t)$ and $\kappa_{T}(t)$ from 4 to 23$\Celsius$ has been reported~\cite{zhouDensitiesIsobaricThermal2015}. The parameters of \eref{eq:temperatureDependenceRho}\eref{eq:temperatureDependenceBeta}\eref{eq:temperatureDependenceKappa} are listed in \tref{tab:params}.
\begin{table}[htpb]
\caption{\label{tab:params}Parameters of \eref{eq:temperatureDependenceRho}\eref{eq:temperatureDependenceBeta}\eref{eq:temperatureDependenceKappa} obtained from ~\cite{zhouDensitiesIsobaricThermal2015}.}
\footnotesize
\begin{indented}
\lineup
\item[]\begin{tabular}{@{}ll}
\br
Parameter&Value \\
\mr
$\rho_0$&854.6$\pm$0.1 kg$\cdot$m$^{-3}$ \\
$\beta_0$&(8.894$\pm$0.094)$\times 10^{-4}~\Celsius^{-1}$ \\
$\kappa_{0}$&(7.738$\pm$0.027)$\times 10^{-4}$ MPa$^{-1}$ \\
$\kappa_t^{\prime}$&(3.227$\pm$0.259)$\times 10^{-6}$ MPa$^{-1}\cdot\Celsius^{-1}$ \\
\br
\end{tabular}
\end{indented}
\end{table}
\normalsize
\begin{figure}[htbp]
\centering
    \includegraphics[width=\columnwidth]{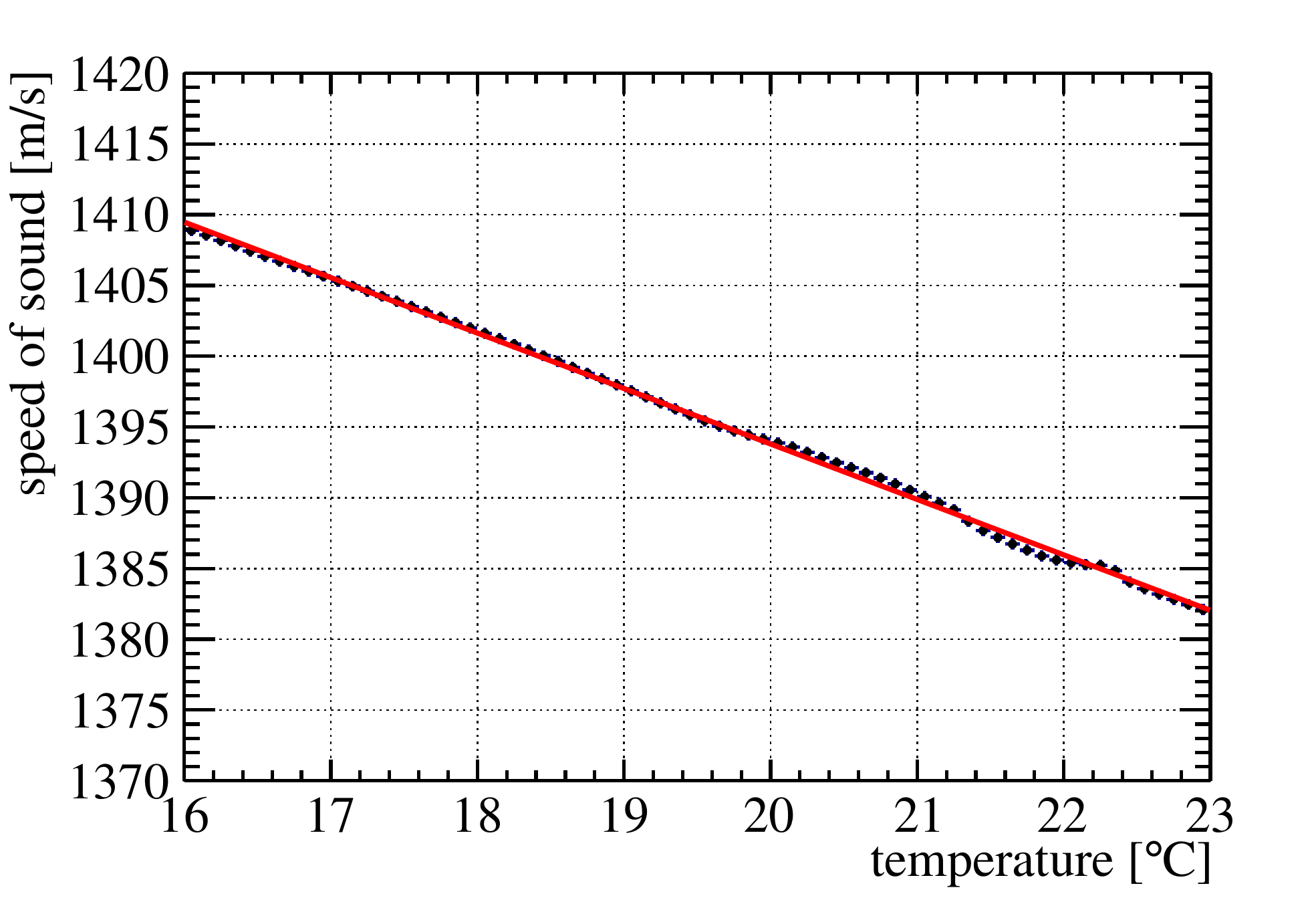}
    \caption{Temperature dependence of speed of sound $v_{\mathrm{sound}}(t)$. Black dots are measured data and the red line is fitting result with \eref{eq:temperatureDependenceVSound}. Fitting result gives $v_{0}$=1382 m$\cdot$s$^{-1}$ and $v_{t}^{\prime}$=$-3.92$ m$\cdot$s$^{-1}\cdot\Celsius^{-1}$}
    \label{fig:vsound}
\end{figure}

\begin{figure}[htbp]
\centering
    \includegraphics[width=\columnwidth]{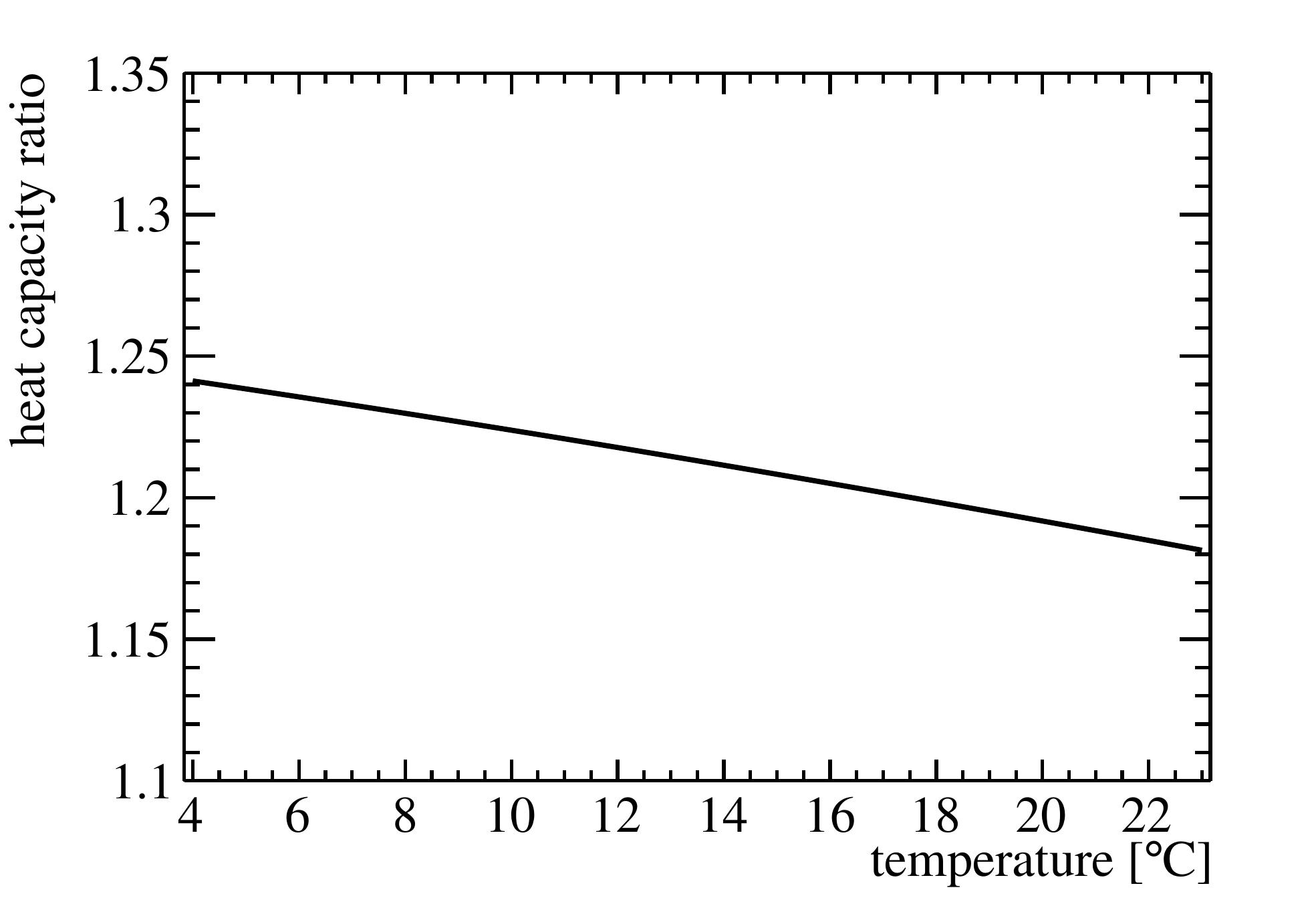}
    \caption{Temperature dependence of heat capacity ratio $\gamma(t)$}
    \label{fig:gamma}
\end{figure}

\begin{figure}[htbp]
\centering
    \includegraphics[width=\columnwidth]{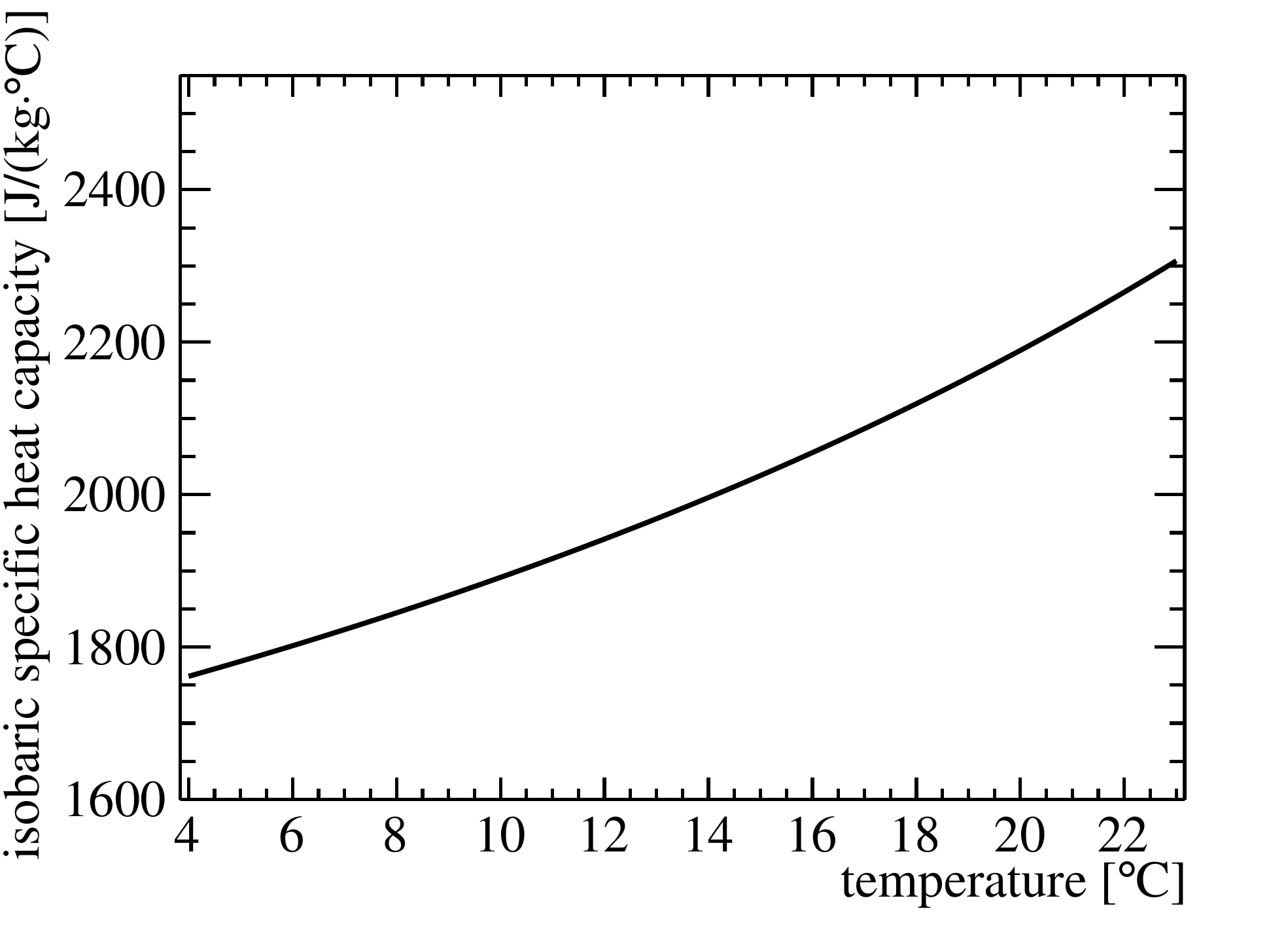}
    \caption{Temperature dependence of isobaric specific heat capacity $c_{P}(t)$}
    \label{fig:cp}
\end{figure}

$v_{\mathrm{sound}}(t)$ from 16$\Celsius$ to 23$\Celsius$ was measured by a commercial sound velocity profiler (Valeport miniSVP). The speed of sound was calculated from the time taken to travel a known distance for a single pulse of sound whose frequency is 2.5 MHz. The measured result is shown as the black points in \fref{fig:vsound}. \Eref{eq:temperatureDependenceVSound} was used to describe the temperature dependence of the speed of sound and extrapolate the lower limit of temperature to 4$\Celsius$. The fitting result is shown as the red line in \fref{fig:vsound} and it gives $v_{0}$=1382 m$\cdot$s$^{-1}$ and $v_{t}^{\prime}$=$-3.92$ m$\cdot$s$^{-1}\cdot\Celsius^{-1}$. There is a 3.4\% difference between $v_{0}$ and \eref{eq:measuredVSound}. \Eref{eq:measuredVSound} is the thermodynamic speed of sound which is the limitation of $v_{0}$ when the frequency of sound approaching zero. Nevertheless, two speed of sound are indistinguishable while the critical frequency is at GHz magnitude which is much larger than 2.5 MHz~\cite{fabelinskiiMolecularScatteringLight1968a}. There is a study shows that alkylbenzenes with numbers of alkyls have different sound speeds~\cite{luningprakDensitiesViscosities2932017}. Since the LAB samples used in two measurements come from different batches, different composition could affect the speed of sound. Assuming that the temperature dependences of sound speed for two batches of LAB are the same, regardless of the absolute value of sound speed. Then the speed of sound in LAB of this study can be written as \eref{eq:temperatureDependenceVSound} with $v_{0}$=1336.34 m$\cdot$s$^{-1}$ and $v_{t}^{\prime}$ remains the same. Therefore, $\gamma(t)$ and $c_{P}(t)$ can be calculated from \eref{eq:temperatureDependenceGamma} and \eref{eq:temperatureDependenceHeatCapacity} respectively. \Fref{fig:gamma} and \fref{fig:cp} show that $\gamma$ decreases and $c_{P}$ increases respectively when temperature increases.


\section{Conclusions and discussions}
We measured the thermal diffusivity and the isobaric specific heat capacity of LAB at about 23$\Celsius$ with 1 atm. The thermal conductivity, heat capacity ratio and the speed of sound were calculated. Results are summarized as following:
\begin{itemize}
    \item $\alpha=(7.245 \pm 0.109)\times10^{-8}~\m^{2}\cdot\s^{-1}$;
    \item $c_{P}=2304\pm23~\J\cdot\Kg^{-1}\cdot\Celsius^{-1};$
    \item $\lambda =(0.1426 \pm0.0026)~\W\cdot\m^{-1}\cdot\Celsius^{-1};$
    \item $\gamma=1.182 \pm 0.005;$
    \item $v_{\mathrm{sound}}=1336.34\pm 3.97~\m\cdot\s^{-1}.$
\end{itemize}
The 3.4\% deviation of speed of sound between two batches of LAB suggests that the final LAB used for the experiment should be measured in situ while the composition is fixed. Based on the assumption that two batches of LAB have the same $v_{t}^{\prime}$, the temperature dependence of heat capacity ratio $\gamma(t)$ and isobaric specific heat capacity $c_{P}(t)$ were estimated.

Considering the nominal value of $c_{P}$, the heat capacity per kiloton LS can be estimated as $2.3\times 10^{9}~\J\cdot\Celsius^{-1}$. It's a large amount of heat which should be taken into account at the LS filling and cooling stage due to the temperature difference of storage and operation. For a typical structure of neutrino experiments, LS is separated from the buffer which is used to shield external radioactivity. For example, the container of LS for JUNO is made of acrylic and water is chosen as the buffer~\cite{anNeutrinoPhysicsJUNO2016}. The conductivity of acrylic is about 0.2 W$\cdot$m$^{-1}\cdot\Celsius^{-1}$ and that of water is about 0.6 W$\cdot$m$^{-1}\cdot\Celsius^{-1}$ which are larger than LAB~\cite{acrylicConducvitity}. Therefore heat is easier transferred from LAB to water, and the residual heat in water could be piled up at the peripheral region of acrylic which would induce the convection of LS. Therefore, it is important to maintain the thermodynamic equilibrium as long as possible which is a challenge for future neutrino experiments.

\ack
This work was supported by the Major Program of the National Natural Science Foundation of China (Grant No. 11390381), the Strategic Priority Research Program of the Chinese Academy of Sciences (Grant No. XDA10010500, XDA10010800), Key Lab of Particle \& Radiation Imaging, Ministry of Education, the Double First Class University Plan of Wuhan University. We thank the helpful discussion with Dr. Jianglai Liu, Dr. Zeyuan Yu and Mr. Xuefeng Ding.

\section*{References}

\begin{thebibliography}{99}

\bibitem{cowanDetectionFreeNeutrino1956}
C.~L. Cowan, F.~Reines, F.~B. Harrison, H.~W. Kruse, and A.~D. McGuire.
\newblock Detection of the {{Free Neutrino}}: A {{Confirmation}}.
\newblock {\em Science}, 124(3212):103--104, July 1956.

\bibitem{kamlandcollaborationFirstResultsKamLAND2003}
{KamLAND Collaboration}, K.~Eguchi, S.~Enomoto, et~al.
\newblock First {{Results}} from {{KamLAND}}: {{Evidence}} for {{Reactor
  Antineutrino Disappearance}}.
\newblock {\em Physical Review Letters}, 90(2):021802, January 2003.

\bibitem{anObservationElectronAntineutrinoDisappearance2012}
F.~P. An, J.~Z. Bai, A.~B. Balantekin, et~al.
\newblock Observation of {{Electron}}-{{Antineutrino Disappearance}} at {{Daya
  Bay}}.
\newblock {\em Physical Review Letters}, 108(17):171803, April 2012.

\bibitem{renocollaborationObservationReactorElectron2012}
{RENO Collaboration}, J.~K. Ahn, S.~Chebotaryov, et~al.
\newblock Observation of {{Reactor Electron Antineutrinos Disappearance}} in
  the {{RENO Experiment}}.
\newblock {\em Physical Review Letters}, 108(19):191802, May 2012.

\bibitem{doublechoozcollaborationIndicationReactorOverline2012}
{Double Chooz Collaboration}, Y.~Abe, C.~Aberle, et~al.
\newblock Indication of {{Reactor}}
  $\overline{\nu}_{e}$
  {{Disappearance}} in the {{Double Chooz Experiment}}.
\newblock {\em Physical Review Letters}, 108(13):131801, March 2012.

\bibitem{andringaCurrentStatusFuture2016}
S.~Andringa, E.~Arushanova, S.~Asahi, et~al.
\newblock Current {{Status}} and {{Future Prospects}} of the {{SNO}}+.
\newblock {\em Advances in High Energy Physics}, 2016.

\bibitem{bravo-bergunoBorexinoThermalMonitoring2018}
D.~{Bravo-Bergu\~no}, R.~Mereu, P.~Cavalcante, et~al.
\newblock The {{Borexino Thermal Monitoring}} \& {{Management System}} and
  simulations of the fluid-dynamics of the {{Borexino}} detector under
  asymmetrical, changing boundary conditions.
\newblock {\em Nuclear Instruments and Methods in Physics Research Section A:
  Accelerators, Spectrometers, Detectors and Associated Equipment}, 885:38--53,
  March 2018.

\bibitem{fischerSearchNeutrinolessDoublebeta2018}
Vincent Fischer.
\newblock Search for neutrinoless double-beta decay with {{SNO}}+.
\newblock {\em arXiv:1809.05986 [hep-ex, physics:physics]}, September 2018.

\bibitem{anNeutrinoPhysicsJUNO2016}
Fengpeng An, Guangpeng An, Qi~An, et~al.
\newblock Neutrino physics with {{JUNO}}.
\newblock {\em Journal of Physics G: Nuclear and Particle Physics},
  43(3):030401, 2016.
  
\bibitem{beacomPhysicsProspectsJinping2017}
John~F. Beacom, Shaomin Chen, Jianping Cheng, et~al.
\newblock Physics prospects of the {{Jinping}} neutrino experiment.
\newblock {\em Chinese Physics C}, 41(2):023002, 2017.

\bibitem{wurmNextgenerationLiquidscintillatorNeutrino2012a}
Michael Wurm, John~F. Beacom, Leonid~B. Bezrukov, et~al.
\newblock The next-generation liquid-scintillator neutrino observatory
  {{LENA}}.
\newblock {\em Astroparticle Physics}, 35(11):685--732, June 2012.

\bibitem{dingNewGadoliniumloadedLiquid2008}
Yayun Ding, Zhiyong Zhang, Jinchang Liu, et~al.
\newblock A new gadolinium-loaded liquid scintillator for reactor neutrino
  detection.
\newblock {\em Nuclear Instruments and Methods in Physics Research Section A:
  Accelerators, Spectrometers, Detectors and Associated Equipment},
  584(1):238--243, January 2008.

\bibitem{wangThermalDiffusivityDiisopropyl2014}
Sheng Wang, Ying Zhang, Maogang He, Shi Zhang, and Xiong Zheng.
\newblock Thermal diffusivity of di-isopropyl ether ({{DIPE}}) in the
  temperature range 298\textendash{{530K}} and pressure up to {{10MPa}} from
  dynamic light scattering ({{DLS}}).
\newblock {\em Fluid Phase Equilibria}, 376:202--209, August 2014.

\bibitem{heMeasurementIsobaricHeat2015}
Maogang He, Chao Su, Xiangyang Liu, Xuetao Qi, Nan Lv. 
\newblock Measurement of isobaric heat capacity of pure water up to supercritical conditions. 
\newblock {\em The Journal of Supercritical Fluids}, 100:1--6, May 2015.


\bibitem{zhouDensitiesIsobaricThermal2015}
X.~Zhou, Q.~M. Zhang, Q.~Liu, et~al.
\newblock Densities, isobaric thermal expansion coefficients and isothermal
  compressibilities of linear alkylbenzene.
\newblock {\em Physica Scripta}, 90(5):055701, 2015.

\bibitem{landauStatisticalPhysicsThird1980}
L.~D. Landau and E.~M. Lifshitz.
\newblock {\em Statistical {{Physics}}, {{Third Edition}}, {{Part}} 1:
  {{Volume}} 5}.
\newblock {Butterworth-Heinemann}, Amsterdam u.a, 3 edition edition, January
  1980.

\bibitem{fabelinskiiMolecularScatteringLight1968a}
I.~L. Fabelinskii.
\newblock {\em Molecular {{Scattering}} of {{Light}}}.
\newblock {Springer US}, 1968.

\bibitem{luningprakDensitiesViscosities2932017}
Dianne~J. Luning~Prak, Peter~J. Luning~Prak, Jim~S. Cowart, and Paul~C.
  Trulove.
\newblock Densities and {{Viscosities}} at 293.15\textendash{}373.15 {{K}},
  {{Speeds}} of {{Sound}} and {{Bulk Moduli}} at 293.15\textendash{}333.15
  {{K}}, {{Surface Tensions}}, and {{Flash Points}} of {{Binary Mixtures}} of
  n-{{Hexadecane}} and {{Alkylbenzenes}} at 0.1 {{MPa}}.
\newblock {\em Journal of Chemical \& Engineering Data}, 62(5):1673--1688, May
  2017.
  
\bibitem{acrylicConducvitity}
{Engineering ToolBox, (2003). Thermal Conductivity of common Materials and Gases. [online] Available at: $https://www.engineeringtoolbox.com/thermal-cond\\uctivity-d\_429.html$ [Accessed 1 Jan. 2019].}

\end{thebibliography}

\end{document}